\documentclass[english,aps,superscriptaddress, twocolumn]{revtex4-1}
\usepackage[T1]{fontenc}
\usepackage[latin9]{inputenc}
\usepackage{amsmath}
\usepackage{amssymb}
\usepackage{graphicx}
\usepackage{esint,booktabs}
\usepackage{multirow}

\makeatletter

\@ifundefined{date}{}{\date{22 May 2017}}
\makeatother

\usepackage{babel}
\begin{document}
	
	\title{Analytical description of photon beam phase spaces in Inverse Compton Scattering sources}
	
	\author{C. Curatolo} 
	\author{I. Drebot} 
	\affiliation{INFN-Milan, via Celoria 16, 20133 Milano, Italy}
	\author{V. Petrillo} 
	\affiliation{INFN-Milan, via Celoria 16, 20133 Milano, Italy}
	\affiliation{Università degli Studi di Milano, via Celoria 16, 20133 Milano, Italy}
	\author{L. Serafini} 
	\affiliation{INFN-Milan, via Celoria 16, 20133 Milano, Italy}
	
\begin{abstract}

We revisit the description of inverse Compton scattering sources and the photon beams generated therein, emphasizing the behavior of their phase space density distributions and how they depend upon those of the two colliding beams of electrons and photons. Main objective is to provide practical formulas for bandwidth, spectral density, brilliance, which are valid in general for any value of the recoil factor, i.e. both in the Thomson regime of negligible electron recoil, and in the deep Compton recoil dominated region, which is of interest for gamma-gamma colliders and Compton Sources for the production of multi-GeV photon beams. We adopt a description based on the center of mass reference system of the electron-photon collision, in order to underline the role of the electron recoil and how it controls the relativistic Doppler/boost effect in various regimes. Using the center of mass reference frame greatly simplifies the treatment, allowing to derive simple formulas expressed in terms of rms momenta of the two colliding beams (emittance, energy spread, etc.) and the collimation angle in the laboratory system. Comparisons with Monte Carlo simulations of inverse Compton scattering in various scenarios are presented, showing very good agreement with the analytical formulas: in particular we find that the bandwidth dependence on the electron beam emittance, of paramount importance in Thomson regime, as it limits the amount of focusing imparted to the electron beam, becomes much less sensitive in deep Compton regime, allowing a stronger focusing of the electron beam to enhance luminosity without loss of mono-chromaticity. A similar effect occurs concerning the bandwidth dependence on the frequency spread of the incident photons: in deep recoil regime the bandwidth comes out to be much less dependent on the frequency spread. The set of formulas here derived are very helpful in designing inverse Compton sources in diverse regimes, giving a quite accurate first estimate in typical operational conditions for number of photons, bandwidth, spectral density and brilliance values - the typical figures of merit of such radiation sources.
\end{abstract}

\maketitle

\section{Introduction}
Inverse Compton Scattering sources (ICSs) are becoming increasingly attractive as radiation sources in photon energy regions either not covered by other high brilliance sources (FEL's, synchrotron light sources) or where compactness becomes an important figure of merit, like for advanced X-ray imaging applications to be implemented in university campus, hospitals, museums, etc., i.e. outside of research centers or large scale laboratories \cite{krafft1}. ICSs are becoming the $\gamma$-ray sources of reference in nuclear photonics, photo-nuclear \cite{ sunwu, eli} and fundamental physics \cite{micieli}, thanks to superior performances in spectral densities achievable. Eventually they will be considered for very high energy photon generation (in the GeV to TeV range) since there are no other competing techniques at present, neither on the horizon, based on artificial tools at this high photon energy \cite{pgamma}. As a consequence, a flourishing of design activities is presently occurring in several laboratories \cite{pog,eggl,bech,ach,du,kuroda,joa,chen,cri,luo} and companies \cite{ruth, loewen, ruth2, ovo}, where ICSs are being conceived, designed and built to enable several domains of applications, and ranging from a few keV photon energy up to GeV's and beyond. Designs of ICSs are carried out considering several diverse schemes, ranging from high gradient room temperature pulsed RF Linacs \cite{star, eli, ref:linac} to CW ERL Super-conducting Linacs \cite{hajima, akagi} or storage rings \cite{thomx, sunwu, higs, park,ale}, as far as the electron beam generation is concerned, and from single pulse J-class amplified laser systems running at $100$ Hz to optical cavities (e.g. Fabry-Perot) running at $100$ MHz acting as photon storage rings for the optical photon beams, not to mention schemes based on FEL's to provide the colliding photon beam \cite{hajima, wu, carlsten}.\\
In order to assess the performances of a specific ICSs under design, detailed simulations of the electron-photon beam collision are typically carried out using Monte Carlo codes \cite{cain,miatesi,luo2} able to model the linear and non-linear electron-photon quantum interaction leading to Compton back-scattering events, taking into account in a complete fashion the space-time propagation of the two colliding beams through the interaction point region, including possible multiple scattering events occurring during the overlap of the two pulses. Only in case of negligible electron recoil, i.e. in the so called Thomson regime typical of low energy X-ray ICSs, classical electromagnetic numerical codes (e.g. TSST \cite{linenonlin}), modelling the equivalent undulator radiation emitted by electrons wiggling in the electromagnetic field of the incoming laser pulse, allow to analyze particular situations such as the use of chirped \cite{powers}, tilted \cite{debus} and twisted \cite{pet} lasers.\\
In the recent past some efforts have been developed to carry out analytical treatments of the beam-beam collision physics, embedding the single electron-photon collision from a quantum point of view within a rms distribution of the scattered photon beam \cite{ale, hartemann2,nn,krafft,vit,ccross,var,pot}, or, within a classical framework, integrating the radiated power in the far field on the distributions of the colliding beams \cite{krafft1, linenonlin, hartemann}. This latter approach suffers from a non-conservation of energy and momentum, due to its lack of describing correctly the electron recoil in the scattering process.
We generalized the former approach to take into account in a complete fashion the recoil effect when averaging over the rms momenta of the two colliding beams, leading in this way to expressions for the bandwidth and spectral densities of the emitted photon beam which are valid for any scattering configuration, with the only restriction of considering relativistic electrons colliding head-on with photons of much smaller energy than electrons (typical of ICSs' operational conditions).\\
As extensively reported in the literature, the emitted photon beam formation is always accomplished by forward collimating the high energy back-scattered photons \cite{eli, cardarelli}, which are emitted all over the solid angle, though with a specific energy-angle correlation such that the most energetic photons propagate around the direction of motion of the electron beam within a small angle O(1/$\gamma$), where $\gamma$ is the relativistic factor of the electron. As illustrated in Section \ref{sec:theory}, carrying-out the kinematics of the electron-photon collision in the center of mass reference system allows to underline that the effective angle of collimation of high energy back-scattered photons is actually 1/$\gamma_{CM}$, where $\gamma_{CM}$ is the center of mass relativistic factor ($\gamma_{CM}$ is always smaller than $\gamma$, approaching $\gamma$ when the electron recoil tends to zero, i.e. in Thomson regime). Using the center of mass reference frame the Doppler frequency enhancement of the scattered radiation can be recovered in the quantum treatment showing that the maximum energy of the scattered photons (the so-called Compton edge) can be expressed as $4\,\gamma_{CM}^2$ times the energy of the incident (optical) photon, therefore generalizing to deep Compton recoil regime the concept of Doppler/boosted frequency enhancement due to the back-scattering process, which is a popular way to describe the behavior of the back-scattered radiation in the Thomson regime.\\
To this purpose Section \ref{sec:theory} is devoted to the illustration of the kinematics in the center of mass reference frame and how a Lorentz transformation to the laboratory frame can bring to a simple exact analytical expression of the back-scattered photon energy at any scattering angle in the laboratory system, as a function of the electron energy $E_e$ and the incident photon energy $E_L$.\\ By recognizing the duality and interplay between the scattering angle and the single electron trajectory crossing angle in the interaction point due to the electron beam emittance, and applying a multi-variate treatment to the photon energy distribution at small angles (i.e. $\gamma_{CM}\theta < 1$ ), we derive a complete expression of the bandwidth of the collimated photon beam within a collimation angle $\theta_{max}$, set by the collimation system, as a function of the incoming beams features, i.e. the electron energy spread and transverse emittance, the laser photon frequency spread and phase front curvature, and the weak non-linear effects represented by the laser parameter $a_0$. By using $\gamma_{CM}$ instead of $\gamma$, the formula fully retrieves the effect of electron recoil to any extent, therefore is applicable to any ICSs regime. As anticipated, the bandwidth dependence on the electron beam emittance is generalized to $2\,(\gamma_{CM} /\gamma)^2\,(\epsilon_n /\sigma_x)^2$, i.e. is basically not dependent on recoil in low recoil regimes (where $\gamma_{CM} = \gamma$) while it scales like the inverse of recoil factor for large recoils (when $\gamma_{CM} \ll \gamma$ ), showing a de-sensitivity of bandwidth from the electron beam emittance, a crucial feature that allows different strategies in large recoil ICSs for beam manipulation in order to maximize luminosity, as discussed further below.\\
Another surprising prediction is that bandwidth is largely independent on frequency spread of the incident photon beam in the deep recoil regime, that brings to a very interesting feature of deep recoil regime: the possibility to generate narrow bandwidth photon beams even using a broad bandwidth incident photon beam, and, at the same time, even over-focusing the electron beam (thanks to the suppressed dependence of bandwidth on electron beam emittance): as reported in Section \ref{sec:sim}, numerical simulations nicely confirm this interesting prediction about a unique feature of deep recoil regime, that eventually leads to possible new options for ICSs designs.\\
These two new findings enhance the analytical description of ICSs with respect to previous works \cite{sunwu, krafft1, krafft, vit, nn} in predicting the behavior of the scattered photon beam phase space in deep recoil, as observed in numerical simulation of Ref. \cite{hajima}. \\
Furthermore, considering the expression of Klein-Nishina differential cross section we derive, through a luminosity treatment of the collision, an approximate expression for the number of photons emitted within a specific collimation angle $\theta_{max}$, under the assumptions of $\gamma_{CM}\,\theta_{max}<1$, therefore also a simple expression for the spectral density (i.e. the number of photons normalized to absolute bandwidth) and brilliance of the emitted radiation beam.\\ Although our analysis is restricted to linear quantum model of the electron-photon collision, as previously said some of the non-linear effects, e.g. those due to high intensity of the incident laser, are integrated in the model to some approximation. 
Extensive comparisons of the analytical formulas derived in Section \ref{sec:theory} versus the results of Monte Carlo simulations are presented in Section \ref{sec:sim}. We selected three different ICSs as paradigmatic of various regimes:
\begin{itemize}
\item[i.]STAR \cite{star}, a typical Thomson Source for X-ray generation in the $20-100$ keV range, devoted to radiological imaging of pre-clinical studies and cultural heritage studies: electron recoil effects are absolutley negligible in this case, where X-ray flux and moderate bandwidth are the key factors (hence maximum luminosity);
\item[ii.]ELI-NP-GBS \cite{eli}, a typical Inverse Compton Source for nuclear photonics and photo-nuclear physics devoted to generate maximum spectral density photon beams in the $1-20$ MeV energy range: here electron recoil is small but non negligible (actually larger than the requested narrow bandwidth);
\item[iii.]XFELO-$\gamma$ \cite{hajima, akagi}, a FEL based Inverse Compton Source for hadronic physics experiments generating up to $7$ GeV photons by back-scattering a $12$ keV FEL beam by a $7$ GeV electron beam circulating in a storage ring: here electron recoil is dominant and strongly affects the bandwidth and intensity of the photon beam. The comparison between analytical predictions and simulation results underline impressively the predicted effect of decreasing the sensitivity of bandwidth to the electron beam emittance by a factor scaling with the inverse of recoil, in such a way that a stronger focusing of the electron beam can be applied without spoiling the bandwidth. As well known this is not possible in low recoil regimes.	
\end{itemize} 
A summary of the formulas in practical forms is presented in the Conclusions, with the aim to offer helpful guidelines to the designers of ICSs in achieving the optimal parameter sets for operation according to the specified requests on photon beam features.

\section{Theory}\label{sec:theory}
\begin{figure}[htbp]\centering
	\includegraphics [scale=0.28] {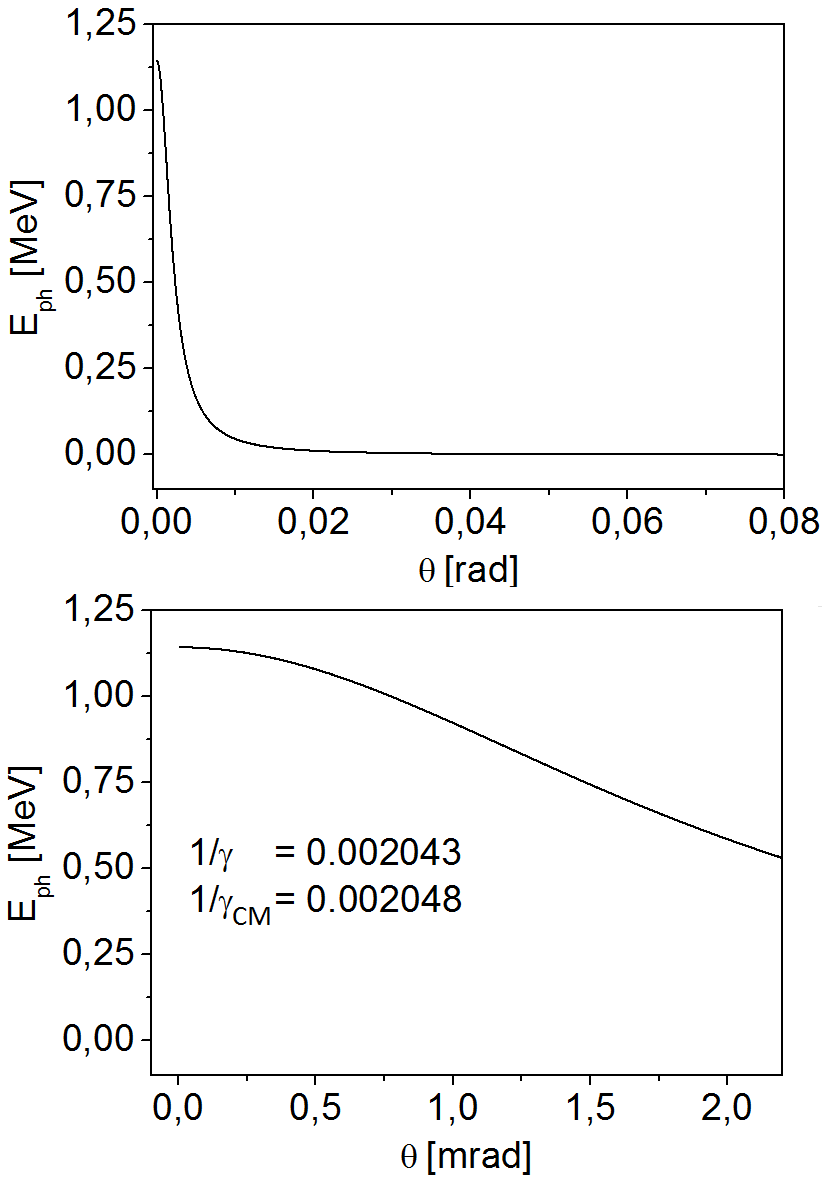}
	\caption{$E_{ph}$ (MeV), energy of the scattered photons in the LAB frame, as a function of $\theta$ for $E_L=1.2$ eV, $\gamma=489.2368\simeq \gamma_{CM}=488.1165$ and $X=0.0046$. Top: $E_{ph}$ (MeV) value up to $\theta=0.08$ rad. Bottom: zoom up to $\theta=220$ mrad.  }\label{es1}
\end{figure}

\begin{figure}[htbp]\centering
	\includegraphics [scale=0.28] {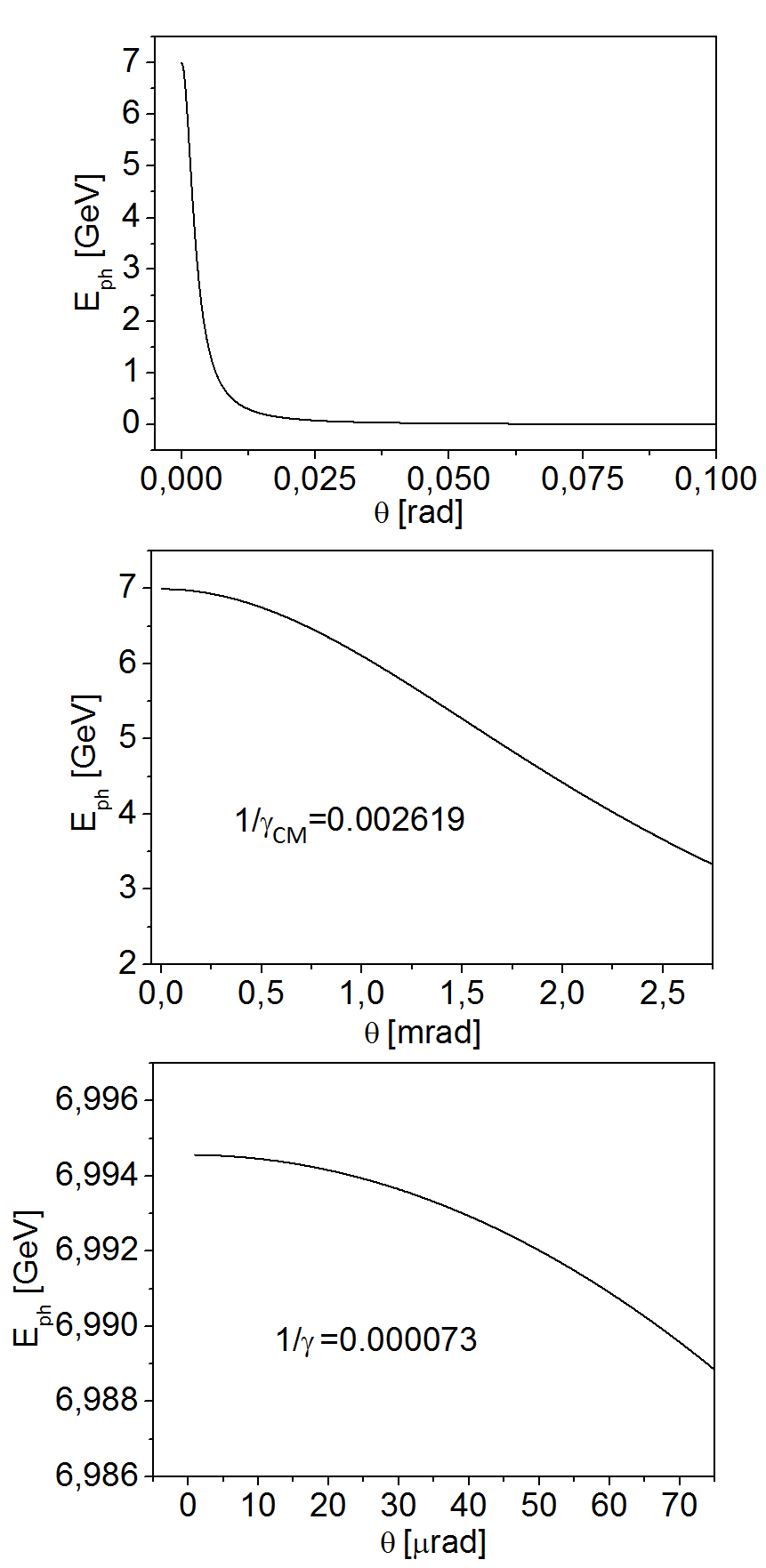}
	\caption{$E_{ph}$ (GeV), energy of the scattered photons in the LAB frame, as a function of $\theta$ for $E_L=12$ keV, $\gamma=13698$, $\gamma_{CM}=381.73$ and $X=1286.9$. Top: $E_{ph}$ (GeV) value up to $\theta=0.1$ rad. Middle: zoom up to $\theta=270$ mrad. Bottom: zoom up to $\theta=80$ $\mu$rad. }\label{es2}
\end{figure}

Let us consider the collision between an electron and a counter-propagating photon of energy respectively $E_{e}$ and $E_{L}$ in the laboratory frame (LAB). We set $c=\hbar=1$. The energy $E'_{L}$ of the colliding photon in the electron rest frame is given by (relativistic Doppler effect)
\begin{equation}
E'_{L}=E_{L}\, \gamma\,(1-\underline{\beta}\cdot\underline{e}_{k})
\end{equation}
where $\underline{\beta}$ is the velocity of the electron, $\underline{e}_{k}$ is the direction of propagation of the photon, $\gamma=E_{e}/M_e$ and $M_e=0.511$ MeV/c$^2$. 
For an ultra-relativistic electron colliding head-on with a photon, the formula simplifies in $E'_{L}\simeq 2\,\gamma\, E_{L}$.
The energy available in the center of mass (CM) of the electron-photon system is 
\begin{equation}
E_{CM}=\sqrt{P^{2}}=\sqrt{2\,E_{e}\,E_{L} -2(\underline{p}_{e}\cdot \underline{k}) +M_e^2}
\end{equation}
where $P=\{E_{e} + E_{L},\, \underline{p}_{e}+\underline{k}\}$ and $\underline{p}_{e}$, $\underline{k}$ the electron and laser photon momenta respectively.
Assuming $E_{e} \gg E_{L}$ and $\gamma \gg1$,

\begin{equation}
\gamma_{CM}=\frac{E_{tot}^{LAB}}{E_{CM}}\simeq\frac{E_{e}+E_{L}}{\sqrt{4\,E_{e}\,E_{L}+M_e^2}}.
\end{equation} 
Once we define the parameter representing the recoil of the electron in the collision as
\begin{equation}
X=\frac{4\,E_e\,E_{L}}{M_e^2}\label{delta},
\end{equation}
we can write $E_{CM}\simeq M_e\sqrt{1+X}$ and $\gamma_{CM}\simeq \gamma/\sqrt{1+X}$.\\ 
\noindent We suppose the electron moves along the positive direction of the $z$ axis in LAB. In the center of mass frame CM the modulus of the momentum of electron and back-scattered photon are
\begin{equation}p^{*}_{e}=E_{ph}^{*}=\frac{E_{CM}^{2}-M_e^2}{2E_{CM}}=\frac{X\,M_e}{2\,\sqrt{1+X}},\label{pion}\end{equation}
showing that there is no threshold for this reaction (as expected, it is a scattering) so that the electron recoil can be arbitrarily small ($*$ denotes the particles' momenta and energies in their CM reference frame). A Lorentz transformation to the LAB gives the energy of the scattered photon as a function of the CM scattering angle $\theta^{*}$ ($\theta^{*}$ calculated with respect to the $z$ axis):
\begin{equation}\begin{split}
E_{ph}=E_{ph}^{*}\,\gamma_{CM}\left(1+\beta_{CM}\cos{\theta^{*}}\right)=\\
4\,E_L\,\gamma^2_{CM}\frac{\left(1+\beta_{CM}\cos{\theta^{*}}\right)}{2},
\end{split}
\label{eph}
\end{equation}
which exhibits 
\begin{equation}\left\{
\begin{aligned}
& E_{ph}^{max}=E_{ph}(\theta^{*}=0)=4\,E_L\,\gamma^2_{CM}=\frac{4\,\gamma^{2}\,E_{L}}{1+X}\\
& E_{ph}^{min}=E_{ph}(\theta^{*}=\pi)=E_{L}
\end{aligned}
\right.
\end{equation}
and we can therefore write
\begin{equation}
X=\frac{E_{ph}^{max}}{E_e-E_{ph}^{max}}.
\end{equation}
Actually, at very small angles in the CM, the corresponding laboratory angle is $\theta=\theta^{*}\sqrt{1+X}/2\,\gamma$, and the photon momentum at small angles around the electron propagation axis (back-scattering close to the Compton edge) is given by
\begin{equation}
E_{ph}=4\,\gamma_{CM}^{2}\,E_{L}\left(1-\gamma_{CM}^{2}\,\theta^{2}\right)
\end{equation}
as well known from the description of the collimated spectral characteristics of Compton sources, that are typically operated with relativistic electrons in very small recoil regime as specified by $X\ll1$.\\ 
Since
\begin{equation}
\tan{\theta}=\left(\frac{\sin{\theta^{*}}}{\gamma_{CM}\left(\beta_{CM}+\cos{\theta^{*}}\right)}\right),
\end{equation}
\begin{equation}
cos{\theta^{*}}=\left\{
\begin{aligned}
\frac{\sqrt{1+\tan^2{\theta}}-\beta_{CM} \, \gamma_{CM}^2 \, \tan^2{\theta}}{1+\gamma_{CM}^2 \, \tan^2{\theta}} \\
\mbox{ \hspace{1cm}if } \theta \le \pi/2 ,\\
\\
\frac{-\sqrt{1+\tan^2{\theta}}-\beta_{CM} \, \gamma_{CM}^2 \, \tan^2{\theta}}{1+\gamma_{CM}^2 \, \tan^2{\theta}} \\
\mbox{ \hspace{1cm}if }  \theta > \pi/2.
\end{aligned}
\right.\label{an}
\end{equation}
\noindent Equations (\ref{eph}) and (\ref{an}) fully specify in a simple analytical form the energy of the scattered photons as a function of $E_e$, $E_L$ and $\theta$.
Examples are reported in Fig. \ref{es1}, \ref{es2} for a small recoil case and a large recoil case respectively.\\

\noindent In the following we analyze the dependence of the emitted photons' relative bandwidth $\Delta E_{ph}/E_{ph}$ from the laser and the electron beam parameters, which are:
$\gamma$ the Lorentz factor, $\Delta \gamma/\gamma$ the relative energy spread, $\epsilon_{n}$ the normalized emittance and $\sigma_{x} $ the rms spot size at interaction point of the electron beam, $\Delta E_L/E_L$ the laser bandwidth, $\lambda_{0}$ the laser wavelength, $w_{0}$ the laser focal spot size, $M^{2}$ the beam quality factor and the laser parameter $a_{0}$. We improve and generalize the formula described in Refs. \cite{vit,eli,linenonlin,nn} by taking into to account the effect given by the electron recoil on the emitted radiation: the use of $\gamma_{CM}$ instead then $\gamma$ extends the validity of the equation to any recoil regime.\\
We define the acceptance angle as
\begin{equation}
\Psi=\gamma_{CM}\theta_{max}
\end{equation}
and the term
\begin{equation}
 \overline{P}=\gamma_{CM}\frac{\sqrt{2}\epsilon_x}{\sigma_x}=\frac{\sqrt{2}\epsilon_n}{\sigma_x\sqrt{1+X}}\label{nu}
\end{equation}
where $\sqrt{2}\epsilon_n/\sigma_x$ represents the normalized rms transverse momentum of the electron beam which coincides with $\overline{P}$ at low recoil. Instead $\overline{P}$ is reduced by a factor $\gamma_{CM}/\gamma\simeq \sqrt{X}$ when the recoil is large. The relative bandwidth of the emitted radiation is given by 
\begin{widetext}
	\begin{equation}
\frac{\Delta E_{ph}}{E_{ph}}\simeq \sqrt{\left[\frac{\Psi^2/\sqrt{12}}{1+\Psi^2}+\frac{\overline{P}^2}{1+\sqrt{12}\,\overline{P}^2}\right]^2+\left[\left(\frac{2+X}{1+X}\right)\frac{\Delta\gamma}{\gamma}\right]^2+\left(\frac{1}{1+X}\frac{\Delta E_L}{E_L}\right)^{2}+
	\left(\frac{M^{2}\lambda_{0}}{2\pi w_{0}}\right)^{4}+\left(\frac{a_{0}^{2}/3}{1+a_{0}^{2}/2}\right)^{2}}\\
\label{f1}
\end{equation} 
\end{widetext}

where in case of a laser Gaussian both in longitudinal and transverse directions
\begin{equation}
a_{0}=  6.8\, \frac{\lambda_0}{w_0}\sqrt{\frac{U_L(J)}{\sigma_t (ps)}} 
\end{equation}
with $U$ the energy of the laser and $\sigma_{t}$ the rms laser pulse length.
\noindent The number of scattered photons per second is given by
\begin{equation}
\mathcal{N}=\mathcal{L}\, \sigma=\frac{N_{e}N_{L}r}{2\pi\left(\sigma_{x}^2+\sigma_L^2\right)} \,\sigma \label{lum}
\end{equation}
where $\mathcal{L}$ is the luminosity,
\begin{equation}\begin{split}
\sigma= \frac{2\pi r_e^2}{X}\left[\frac{1}{2}+\frac{8}{X}-\frac{1}{2(1+X)^2}+\right.\\
\left.\left(1-\frac{4}{X}-\frac{8}{X^2}\right)\log(1+X)\right] 
\end{split}
\end{equation}
is the total unpolarized Compton cross section \cite{landau}, $N_{e}, N_L$ are the number of incoming electrons and photons, $r$ is the repetition rate of the collisions, and $\sigma_{x}$, $\sigma_L=w_0/2$ are the rms spot size radius at the IP of the electron and photon beams respectively. The value of $\sigma$ varies between the classical limit $X\to 0$ and the ultra-relativistic limit $X\to \infty$ as presented in eq. (\ref{lim}) where $\sigma_T=0.67$ barn represents the total Thomson cross section \cite{jackson}.\\
\begin{equation}\left\{
\begin{aligned}
&\lim_{X\to 0}\sigma=\frac{8\pi r_e^2}{3}(1-X)=\sigma_T(1-X) \\
&\lim_{X\to \infty}\sigma=\frac{2\pi r_e^2}{X}\left(\log{X}+\frac{1}{2}\right)
\end{aligned}
\right.\label{lim}
\end{equation}
In practical units, 
\begin{equation}
\mathcal{N}=4.2\cdot 10^8\frac{ \sigma\, U_L(J) \, Q(pC) \, r}{\sigma_T \, E_{L}(eV)\left(\sigma_{x}^2(\mu m)+\sigma_L^2(\mu m)\right)}\label{n}.
\end{equation}
\noindent By using the Compton differential cross section \cite{landau} in the approximation $\Psi <1$, we obtain the analytical expression to estimate $\mathcal{N}^{\Psi}$, the number of photons in acceptance angle $\Psi$, and the spectral density $S$: 
\begin{equation}\begin{split}
\mathcal{N}^{\Psi}=6.25\cdot 10^8 \frac{ U_L(J) \, Q(pC) \, r}{E_{L}(eV)\left(\sigma_{x}^2(\mu m)+\sigma_L^2(\mu m)\right)}\cdot\\
\frac{\left(1+\sqrt[3]{X}\Psi^2/3\right)\Psi^2}{\left(1+(1+X/2)\Psi^2\right)(1+ \Psi^2)},
\end{split}
\label{nbw}
\end{equation}
\begin{equation}
S=\frac{\mathcal{N}^{\Psi}}{\sqrt{2\,\pi}\, 4\,E_L\gamma_{CM}^2 \,\frac{\Delta E_{ph}}{E_{ph}}}.\label{s}
\end{equation}
\noindent The rms source spot size is
\begin{equation}
\sigma_s=\frac{\sigma_x \, \sigma_L}{\sqrt{\sigma_x^2 + \sigma_L^2}}
\end{equation}
and the emittance of the emitted radiation is
\begin{equation}
\epsilon_{\gamma}=\sigma_s\,\frac{\theta_{max}}{\sqrt[4]{12}\,\sqrt[9]{1+X}}\label{ag}.
\end{equation}
\noindent The peak brilliance is defined as
\begin{equation}
B^{peak}=\frac{ \mathcal{N}^{\Psi}} {(2\pi)^{3}\,\epsilon_{\gamma}^2\,\sigma_{t}^{\gamma}\frac{\Delta E_{ph}}{E_{ph}}\, [0.1\%] \, r}\label{b}
\end{equation}
with  $\sigma_{t}^{\gamma}$ the rms duration value of the emitted $\gamma$ photons.\\
The average brilliance on one second is instead given by
\begin{equation}
B^{ave}=\frac{\mathcal{N}^{\Psi}}{(2\pi)^{\frac{5}{2}}\, \epsilon_{\gamma}^2\, \frac{\Delta E_{ph}}{E_{ph}}\, [0.1\%]}.\label{ba}
\end{equation}
 
\section{Simulations}\label{sec:sim}

We will benchmark some of the formulas in the previous section against the simulated values obtained in STAR, ELI-NP-GBS and XFELO-$\gamma$ cases. The simulations have been performed by means of the Monte Carlo codes CAIN and CMCC \cite{cain, miatesi}.\\
From the simulations data, $\Delta E_{ph}/E_{ph}$ is calculated as the rms value of the distributions divided by its mean value:
\begin{equation*}\begin{split}
&\frac{\Delta E_{ph}}{E_{ph}}=\frac{\sqrt{\langle E_{ph}^2\rangle-\langle E_{ph}\rangle^2}}{\langle E_{ph}\rangle}\\
\langle E_{ph}\rangle=\frac{1}{N}&\sum_{i=1,N} {E_{ph}}_i \hspace{1cm}
\langle E_{ph}^2\rangle=\frac{1}{N}\sum_{i=1,N}{E_{ph}}_i^2
\end{split}
\end{equation*}
\noindent where ${E_{ph}}_i$ is the energy of the i-th emitted photon and $N$ the number of photons in the considered set.\\
In the following examples we always consider $\mathcal{N}$ and $\mathcal{N}^{\Psi}$ (eqs. (\ref{n}), (\ref{nbw})) per shot, i.e. $r=1$ and the electron beam rms length equal to the laser pulse one $\sigma_z=c\,\sigma_t$. Furthermore the collisions are perfectly head-on and the beam diffraction throughout the interaction region is negligible so that eq. (\ref{lum}) is applicable because the luminosity is not spoiled by hour-glass effects. 

\subsection{STAR}
\begin{figure}[htbp]\centering
	\includegraphics [scale=0.28] {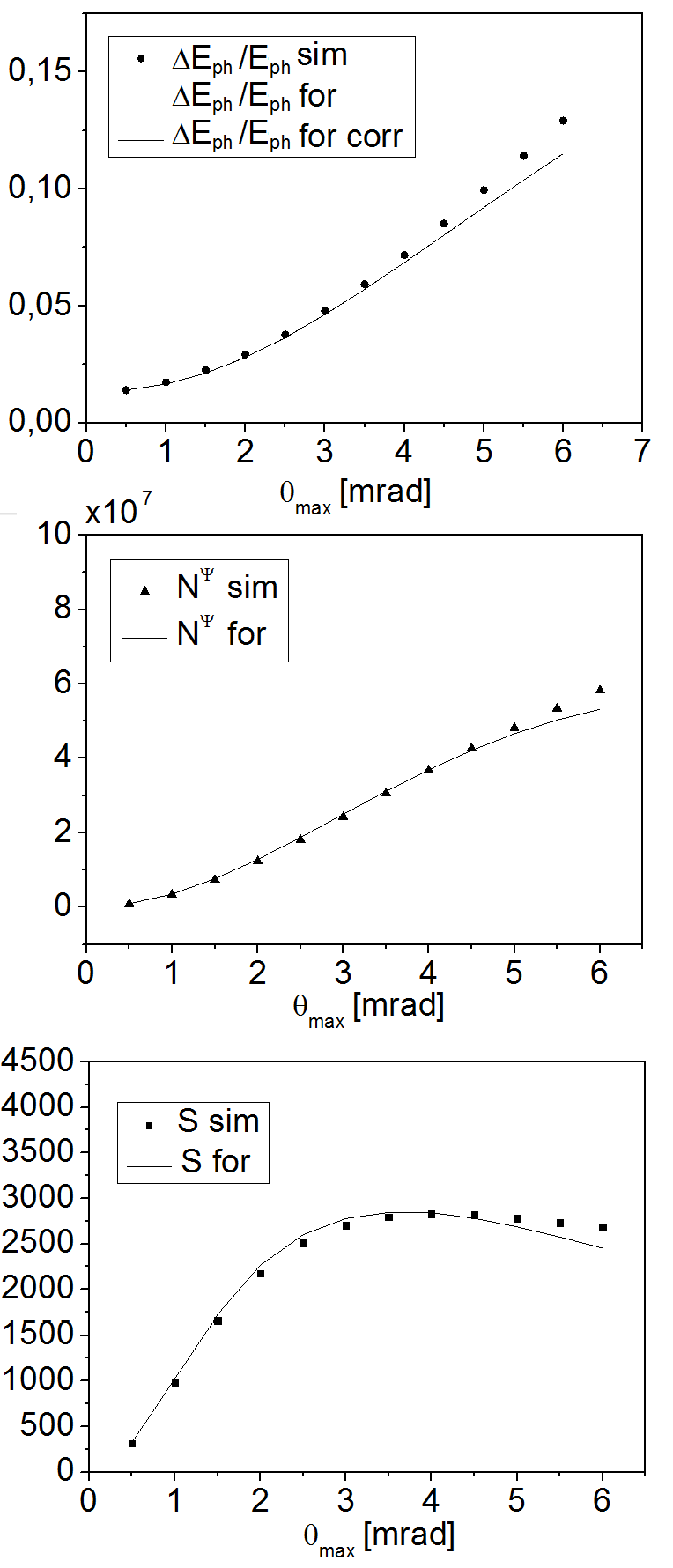}
	\caption{Case A: $\Delta E_{ph}/E_{ph}$ value from CAIN simulation (circles) vs formula (\ref{f1}) without (dotted line) and with (solid line) $X$ correction. Number of photons in $\theta_{max}$: simulated values (triangles) vs formula (\ref{nbw}) (solid line). Spectral density per shot: simulated values (squares) vs formula (\ref{s}) (solid line).}\label{caseA}
\end{figure}

\begin{figure}[htbp]\centering
	\includegraphics [scale=0.28] {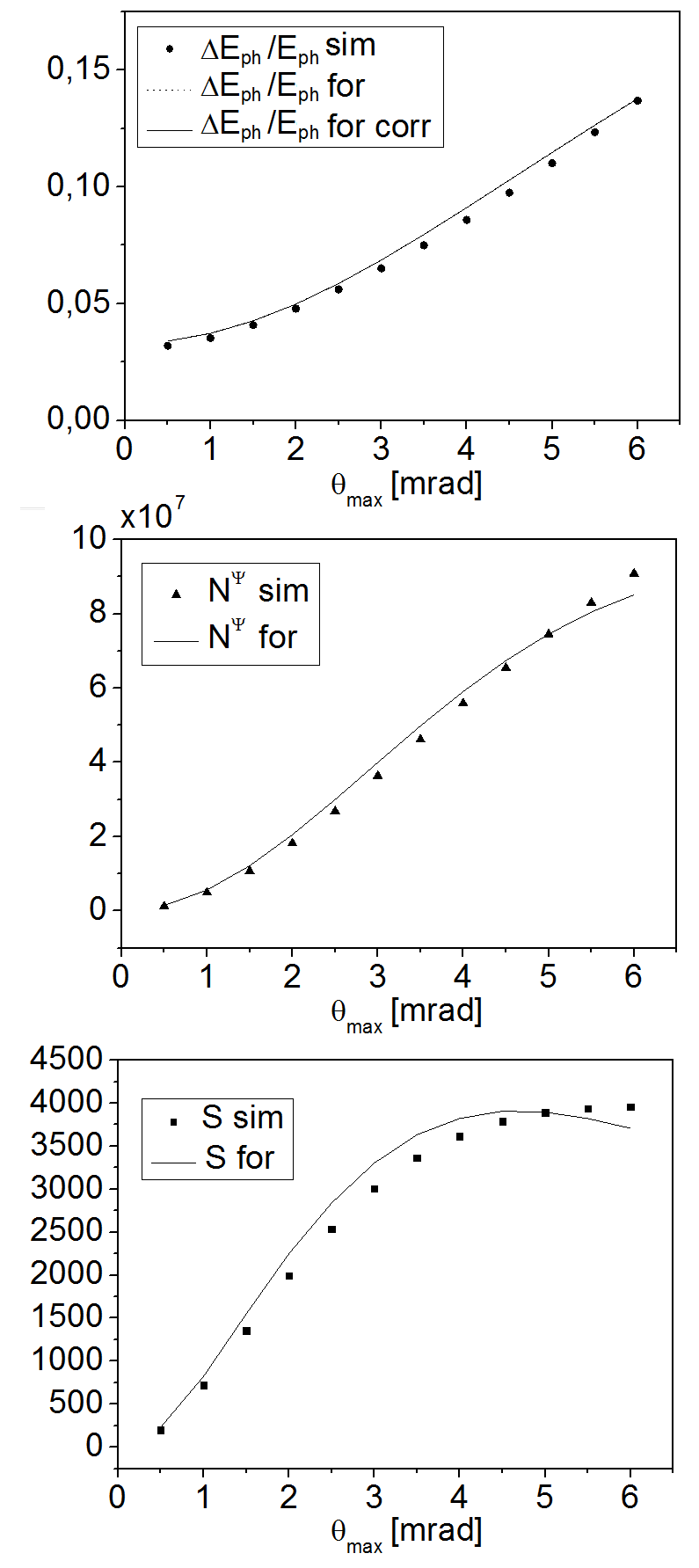}
	\caption{Case B: $\Delta E_{ph}/E_{ph}$ value from CAIN simulation (circles) vs formula (\ref{f1}) without (dotted line) and with (solid line) $X$ correction. Number of photons in $\theta_{max}$: simulated values (triangles) vs formula (\ref{nbw}) (solid line). Spectral density per shot: simulated values (squares) vs formula (\ref{s}) (solid line).}\label{caseB}
\end{figure}

The Southern European Thomson source for Applied Research (STAR), under construction at the University of Calabria, is a typical example of Thomson source. We report in Table \ref{star} the interaction parameters: the very low $X$ value enables a classical approach to this source study. Two different focusing of the electron beam have been considered: in both cases the impact of the recoil parameter on the bandwidth value is negligible (see Figs. (\ref{caseA}), (\ref{caseB})). In case B, $\Delta E_{ph}/E_{ph}$ value is higher than in case A at small collimation angles ($\bar{P}$ is doubled) and also the number of photons $\mathcal{N}^{\Psi}$ and the spectral density $S$ are higher in case B at any $\theta_{max}$.\\

\begin{figure}[htbp]\centering
	\includegraphics [scale=0.28] {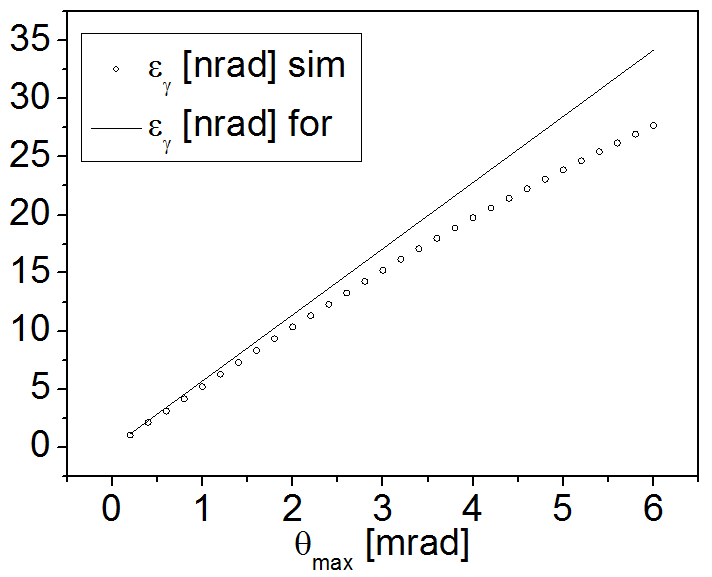}
	\caption{Case A: $\epsilon_{\gamma}$ value from CAIN simulation (circles) vs formula (\ref{ag}) (solid line) (nrad) as a function of $\theta_{max}$ (mrad).}\label{emittgammastar}
\end{figure}

\begin{table}[htbp]
	\centering	
	\caption{Interaction parameters for STAR. $\sigma_L=15$ $\mu$m, $\sigma_t=1$ ps.}\label{star}
	\begin{tabular}{ccccccccc} \\
		\toprule
		Case & Q & $E_{e}$ &   $\Delta\gamma/\gamma$& $\epsilon_n$ & $\sigma _{p_x}$ &$\sigma_x$ & $\lambda_0$ & U$_L$ \\ 
		 & (nC) &  (MeV) & ($10^{-3}$) & ($\mu$m\ rad) & (keV) & ($\mu$m) &  ($\mu$m)  &  (J) \\ 			
			\midrule
			A &	1 & $65$ & $5$  & 1  & 34  & 15  & $1$  & 0.2  \\
			B &	1 & $65$ & $5$  & 1  & 68  & 7.5 &  $1$ & 0.2  \\ 
			\midrule
			\multicolumn{9}{c}{$X=0.00123$}\\
			\bottomrule 
	\end{tabular}		
\end{table}
\noindent Fig.(\ref{emittgammastar}) exhibits a very good agreement between formula (\ref{ag}) and the value of the emitted photon beam emittance given by the CAIN simulations.

\subsection{ELI-NP-GBS}
\begin{figure}[htbp]\centering
	\includegraphics [scale=0.28] {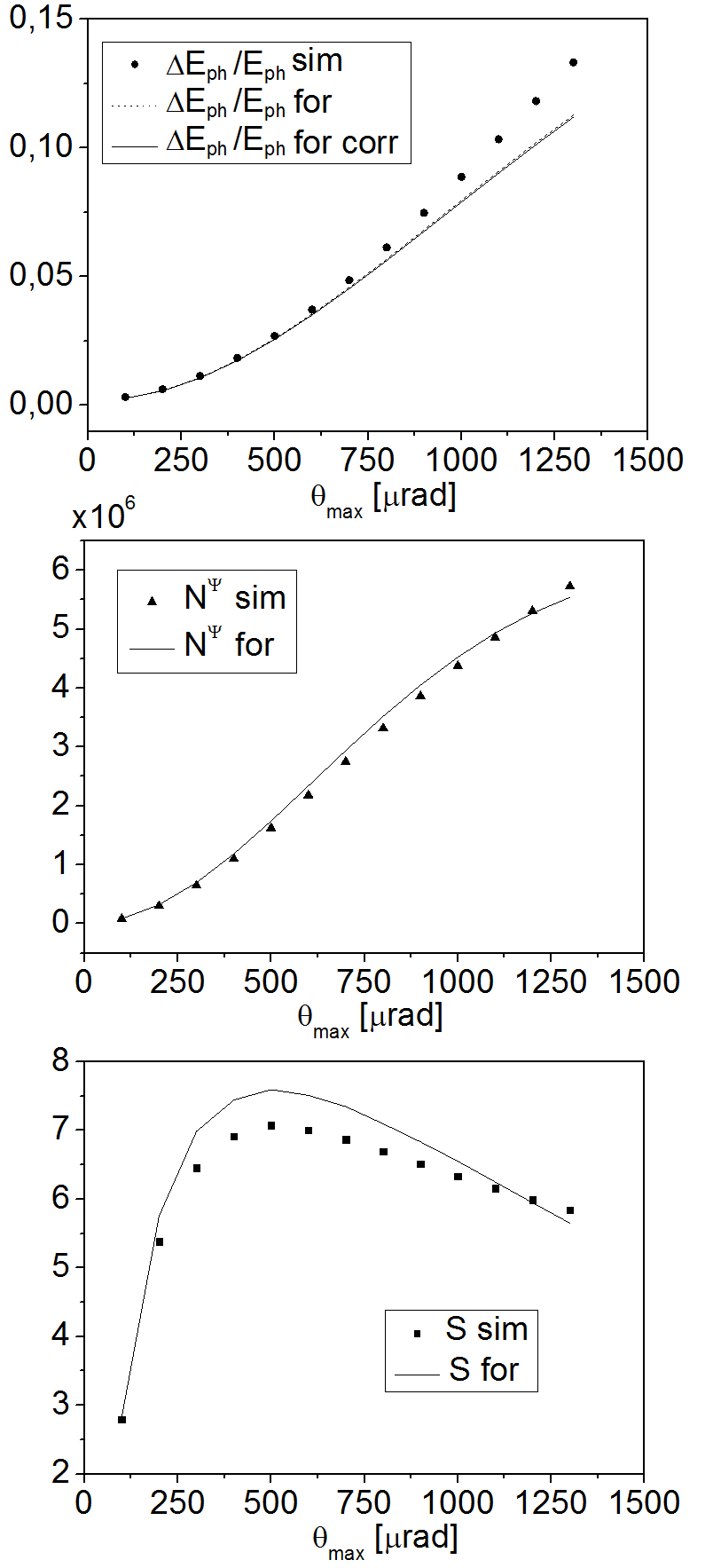}
	\caption{Case C: $\Delta E_{ph}/E_{ph}$ value from CAIN simulation (circles) vs formula (\ref{f1}) without (dotted line) and with (solid line) $X$ correction. Number of photons in $\theta_{max}$: simulated values (triangles) vs formula (\ref{nbw}) (solid line). Spectral density per shot: simulated values (squares) vs formula (\ref{s}) (solid line).}\label{caseC}
\end{figure}

\begin{figure}[htbp]\centering
	\includegraphics [scale=0.28] {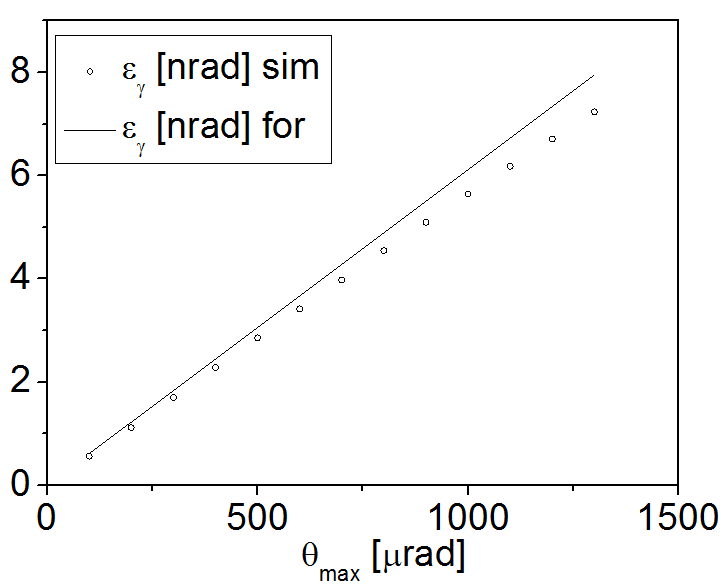}
	\caption{Case C: $\epsilon_{\gamma}$ value from CAIN simulation (circles) vs formula (\ref{ag}) (solid line) (nrad) as a function of $\theta_{max}$ (mrad).}\label{emittgammaeli}
\end{figure}

The Extreme Light Infrastructure Nuclear Physics Gamma Beam System is a linear machine based on the collision of an intense high power Yb:Yag J-class laser and a high brightness electron beam with a tunable energy up to $750$ MeV. The main specifications of the Compton Source are: photon energy tunable in the  $0.2-19.5$ MeV energy range, rms relative bandwidth smaller than $0.5\%$ and spectral density lager than $5 \cdot 10^3$ photons/s$\cdot$eV, with source spot sizes smaller than $100$ $\mu$m and linear polarization of the gamma-ray beam larger than $95\%$. Moreover the peak brilliance of the $\gamma$ beam is expected to be larger than $10^{19}$ photons/(s$\cdot$mm$^2\cdot$mrad$^2\cdot$0.1\%). The recoil parameter is quite low, nevertheless its impact is not negligible since the bandwidth value request is highly demanding. Fig. \ref{caseC} suggests a $\theta_{max}\ll 1/\gamma_{CM}=1220$ $\mu$rad to obtain $\Delta E_{ph}/E_{ph}=0.005$ and Fig.\ref{emittgammaeli} shows the S peak around an angle of about $400$ $\mu$rad.
\begin{table}[htbp]
	\centering	
	\caption{Interaction parameters for ELI-NP-GBS. $\sigma_L=14$ $\mu$m, $\sigma_t=1.5$ ps.}\label{eli}
	\begin{tabular}{ccccccccc} \\
		\toprule
		Case & Q & $E_{e}$  &   $\Delta\gamma/\gamma$& $\epsilon_n$ & $\sigma _{p_x}$ & $\sigma_x$ &  $\lambda_0$  & U$_L$ \\
		& (pC) &  (MeV) & ($10^{-4}$) & ($\mu$m\ rad) & (keV) & ($\mu$m) &  (nm)  &  (J) \\  			
		\midrule
		C & $250$  & $311.65$ & $7 $ & $0.5$  & $13.2$ & $19.6$  & $515$ & $0.2$ \\
		\midrule
		\multicolumn{9}{c}{$X=0.011$}\\
		\bottomrule 
	\end{tabular}		
\end{table}

\subsection{XFELO-$\gamma$}
\begin{figure}[htbp]\centering
		\includegraphics [scale=0.3] {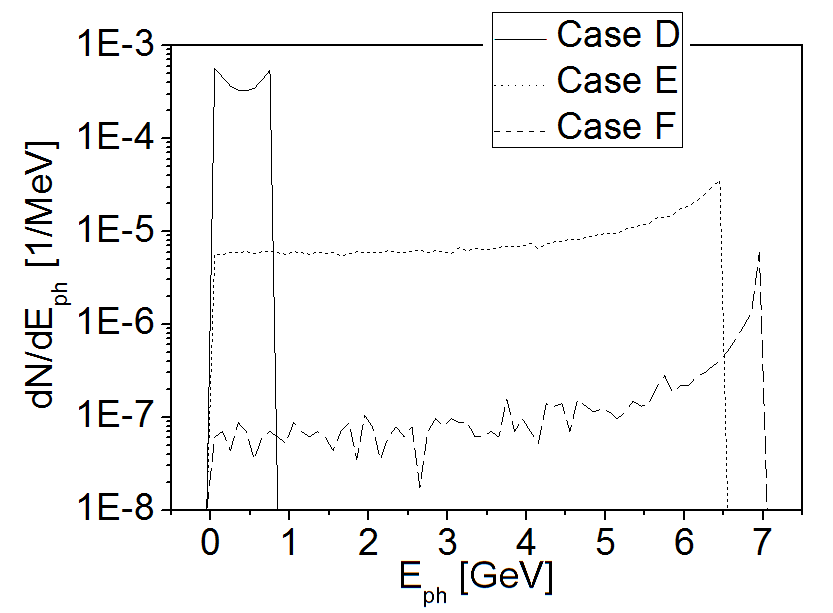}
		\caption{Spectrum of the emitted photons for the three XFELO-$\gamma$ cases, CMCC simulations.}\label{fig:4}
\end{figure}

\begin{figure*}[htbp]\centering
	\includegraphics [width=0.95\textwidth] {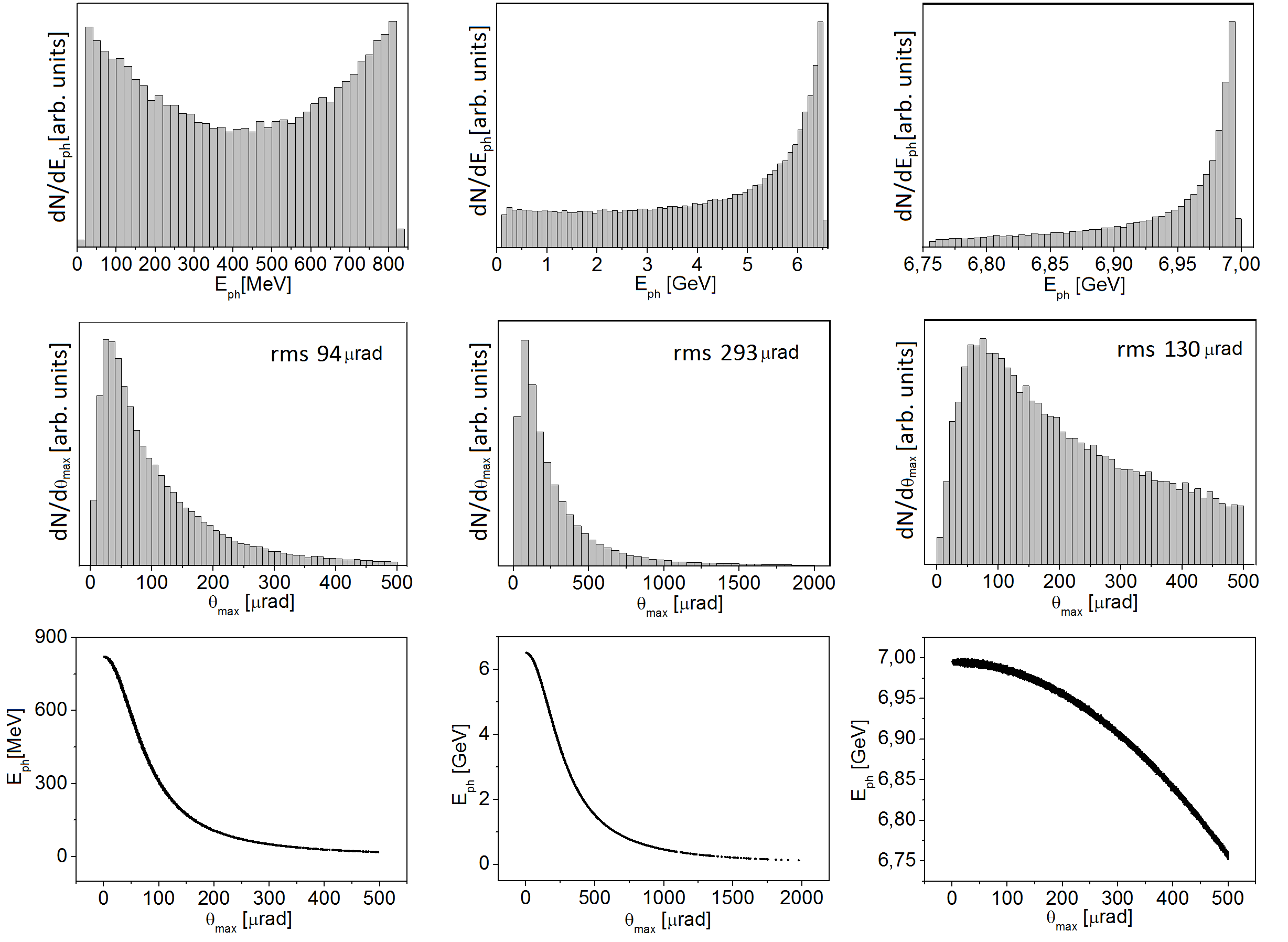}
	\caption{Emitted photons features, CAIN simulations. First line energy distribution, second line angle distribution, third line energy as a function of the angle. Case D left column, case E middle column, case F right column. }\label{caseD}
\end{figure*}

The XFELO-$\gamma$ design described in \cite{hajima} considers the collision between a $7$ GeV electron beam and three possible  photon beam energies covering a wide range of recoil regimes: a $1.239$ eV photon beam leads to a low recoil parameter $X=0.1328$, an intermediate case at $123.9$ eV sets $X=13.28$ and a very high recoil regime is reached at $12.39$ keV. The emitted radiation spectra are reported for all the cases in Fig. \ref{fig:4} and they are analyzed in details in Fig. \ref{caseD}.\\ 

\begin{figure*}[htbp]\centering
	\includegraphics [width=0.95\textwidth] {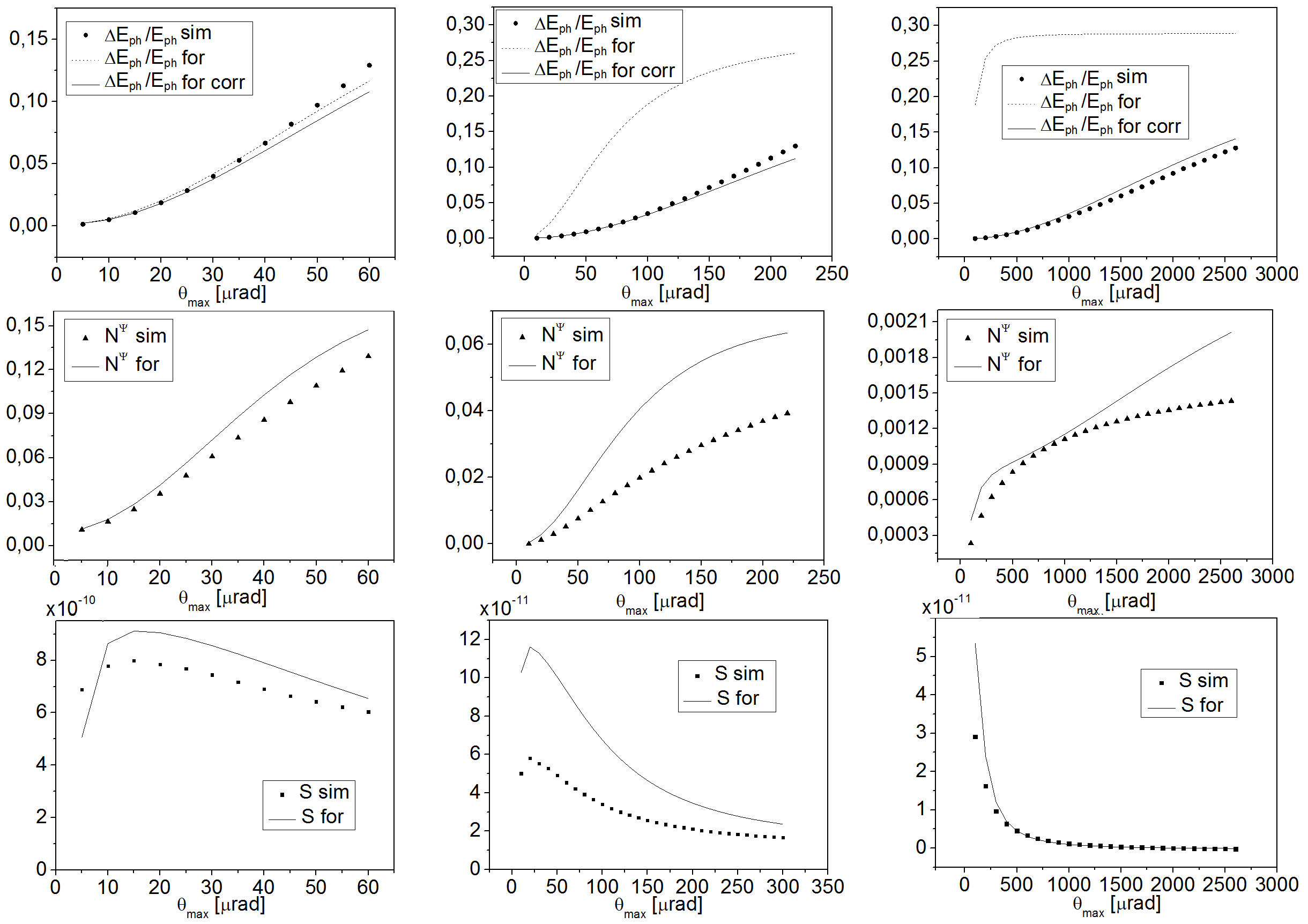}
	\caption{$\Delta E_{ph}/E_{ph}$ value from CAIN simulation (circles) vs formula (\ref{f1}) without (dotted line) and with (solid line) $X$ correction. Number of photons in $\theta_{max}$: simulated values (triangles) vs formula (\ref{nbw}) (solid line). Spectral density per shot: simulated values (squares) vs formula (\ref{s}) (solid line). Case D left column, case E middle column, case F right column.}\label{low}
\end{figure*}

\begin{figure}[htbp]\centering
	\includegraphics [scale=0.28] {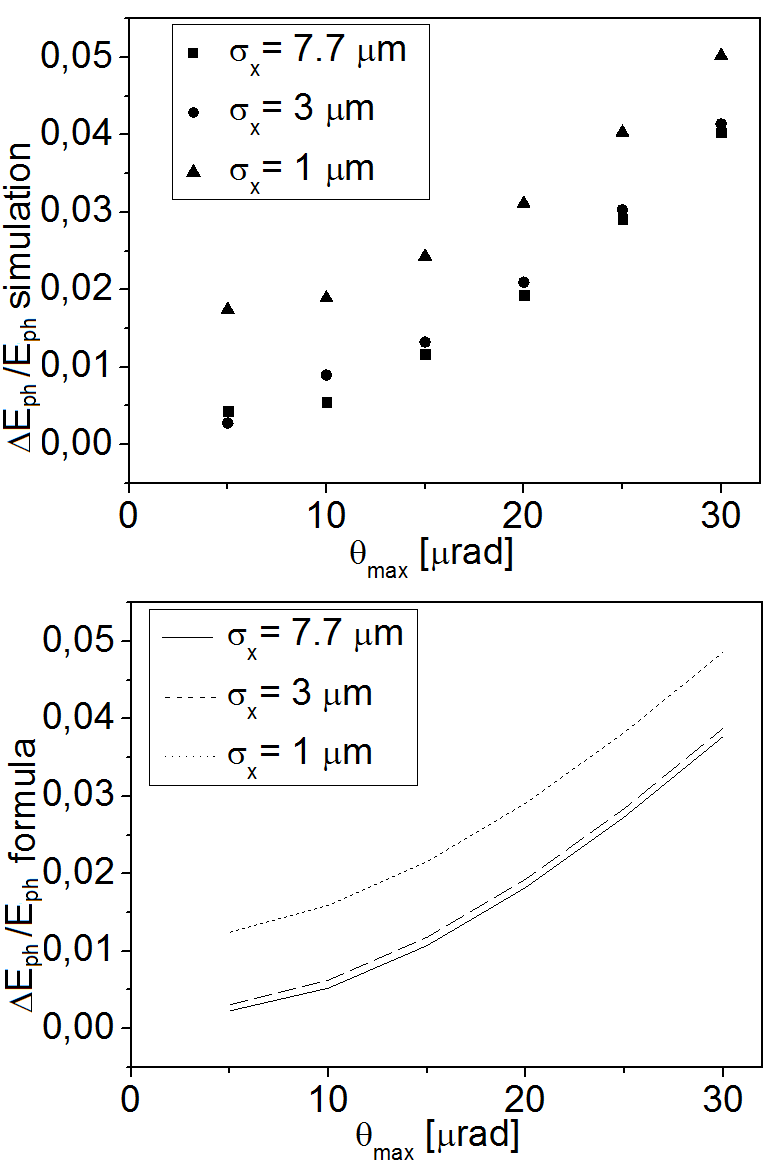}
	\caption{Case D for different $\sigma_x$ values and constant $\epsilon_n$.}\label{foc}
\end{figure}

\begin{figure}[htbp]\centering
	\includegraphics [scale=0.28] {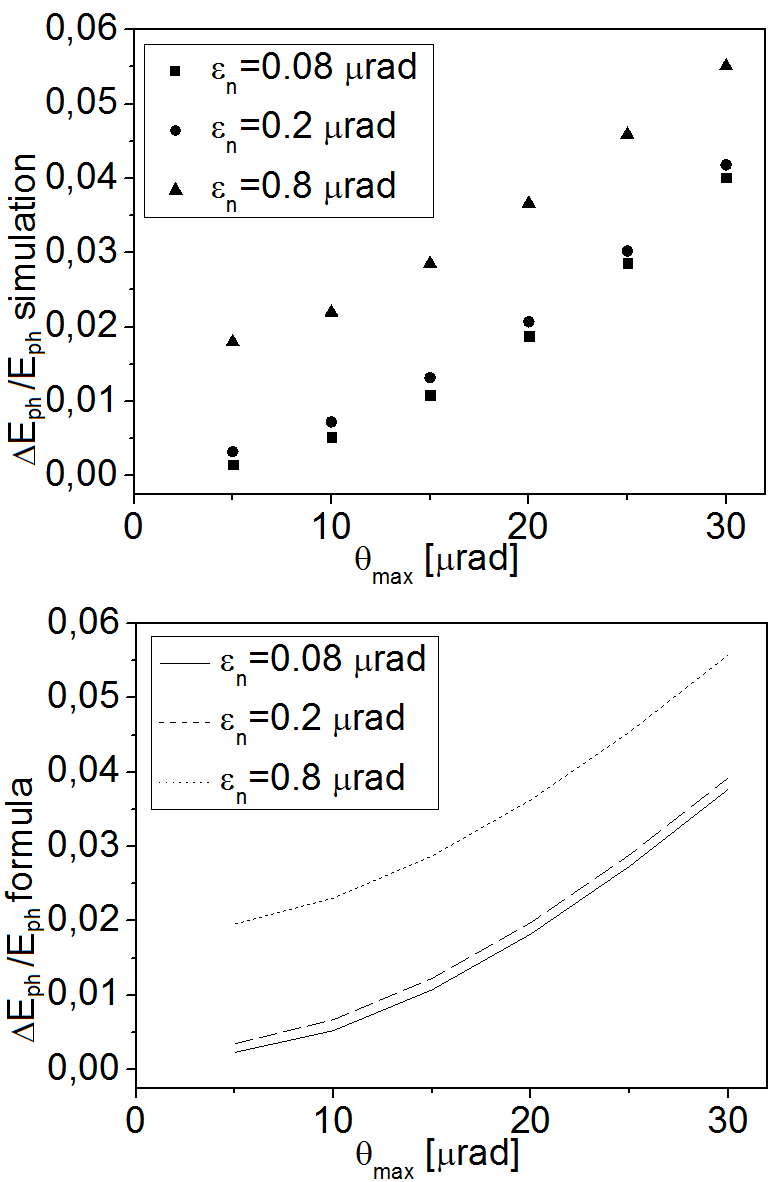}
	\caption{Case D for different $\epsilon_n$ values.}\label{emit}
\end{figure}

\noindent The low $X$ case (D) exhibits the typical Thomson spectrum shape, $\gamma\simeq\gamma_{CM}$ and half of the emitted photons are contained in the small $1/\gamma_{CM}\simeq 1/\gamma$ angle. Fig. \ref{low} shows a good agreement between the low recoil case and the classical treatment ($X=0$), Fig. \ref{foc} and \ref{emit} report a very good agreement between simulations and formulas concerning the bandwidth calculation at small angles for different $\sigma_x$ and emittance values.\\

\begin{table}[htbp]
	\centering	
	\caption{Interaction parameters XFELO-$\gamma$. Gaussian distribution for position and momentum of the electrons.  $N_L=2\cdot 10^{10}$, $\sigma_L=8.9$ $\mu$m and  $\sigma_t=0.85$ ps.}
	\begin{tabular}{cccccccccc} \\
		\toprule
		Case & Q & $E_{e}$ &   $\Delta\gamma/\gamma$& $\epsilon_n$ & $\sigma _{p_x}$ &$\sigma_x$ & $\lambda_0$  & U$_L$\\
		& (pC) &  (GeV) & ($10^{-4}$) & ($\mu$m\ rad) & (keV) & ($\mu$m) &  (nm)  &  ($\mu$J)\\
		\midrule
		D & $40$ & $7$ & $2$ & $0.082$ & $5.4$ & $7.7$  & $1000$ & $0.00397$\\
		E & $40$ & $7$ & $2$ & $0.082$ & $5.4$ & $7.7$  & $10$   & $0.397$  \\
		F &$40$  & $7$ & $2$ & $0.082$ & $5.4$ & $7.7$  & $0.1$  & $39.7$    \\
		\bottomrule 
	\end{tabular}		
\end{table}

\begin{table}[htbp]
	\centering	
	\caption{Interaction parameters XFELO-$\gamma$.}
	\begin{tabular}{cccccccc} \\
		\toprule
		Case & $X$	& $E_{CM}$  & $\gamma$ & $\gamma_{CM}$ & 1/$\gamma$ & 1/$\gamma_{CM}$ &$\mathcal{N}$\\
		& & (MeV)& & & ($\mu$rad) & ($\mu$rad)& \\ 			
		\midrule
		D & 0.1328 & 0.542 & 13700 &  12915 & 73  & 77 & 0.34\\
		E & 13.28 & 1.93 & 13700 &  3626.94 & 73  & 275 & $6.3 \cdot 10^{-2}$\\
		F & 1328 & 18.36 & 13700 & 375.73 & 73 & 2660 & $1.7 \cdot 10^{-3}$\\ 
		\bottomrule 
	\end{tabular}	
\end{table}

\noindent The spectrum in the medium recoil case E (Fig. \ref{caseD}) shows an appreciable asymmetry towards the high energies given by the strong asymmetry of the differential cross section. In this intermediate case the $X$ contribution to the bandwidth value calculation is not negligible as shown in Fig. \ref{low}.\\

\begin{figure}[htbp]\centering
	\includegraphics [scale=0.3] {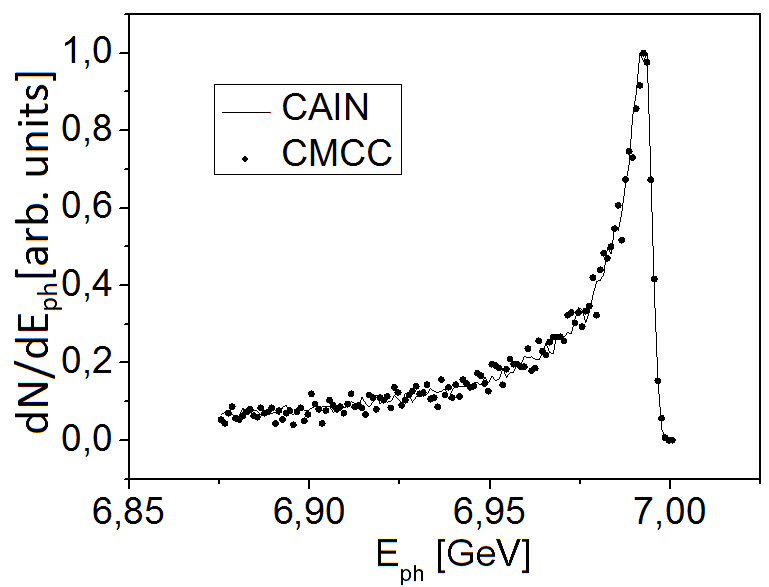}
	\caption{Case F: emitted photons spectrum. CAIN vs CMCC simulations.}\label{highvs}
\end{figure}

\begin{figure}[htbp]\centering
	\includegraphics [scale=0.28] {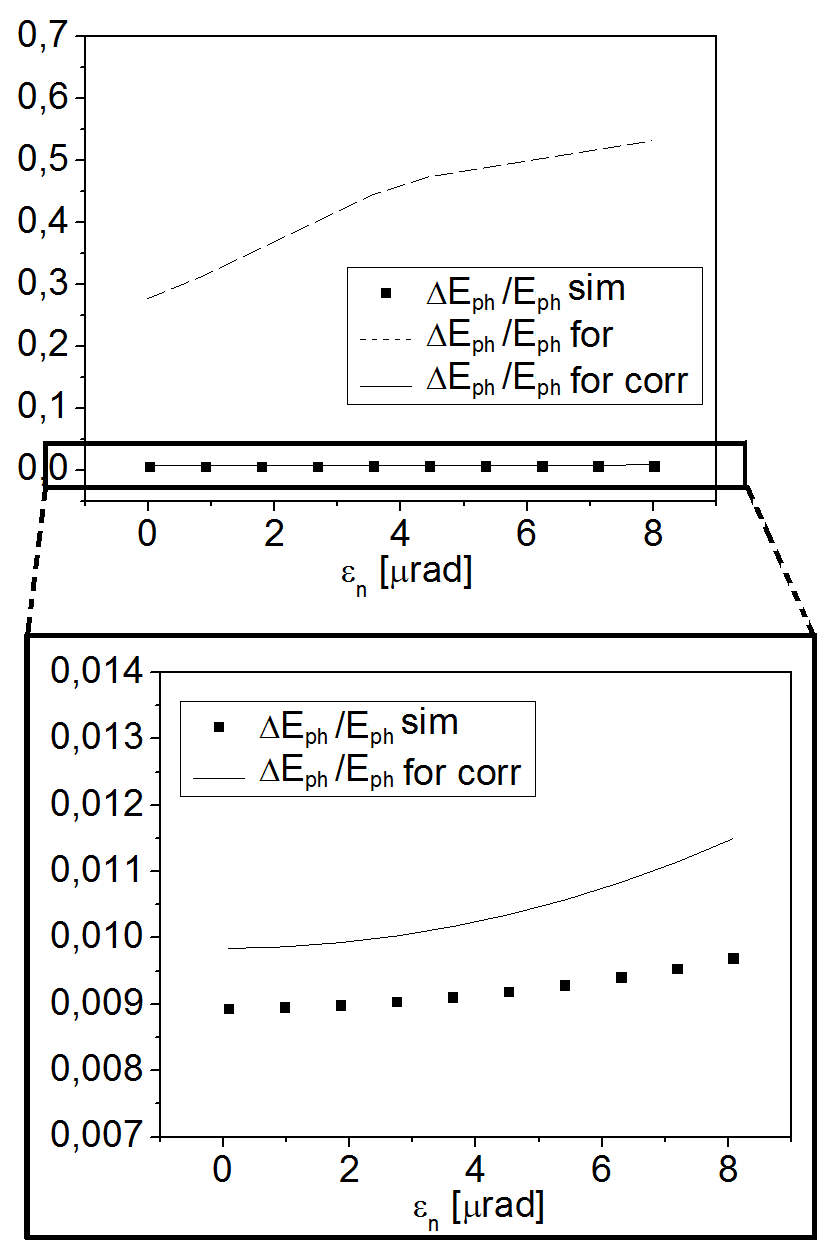}
	\caption{Case F: $\Delta E_{ph}/E_{ph}$ value from CAIN simulation (squares) vs formula (\ref{f1}) without (dashed line) and with (solid line) $X$ correction for different $\epsilon_n$ ($\mu$m\ rad) values calculated at $\theta=500$ $\mu$rad.}\label{emitthigh}
\end{figure}

\begin{figure}[htbp]\centering
	\includegraphics [scale=0.28] {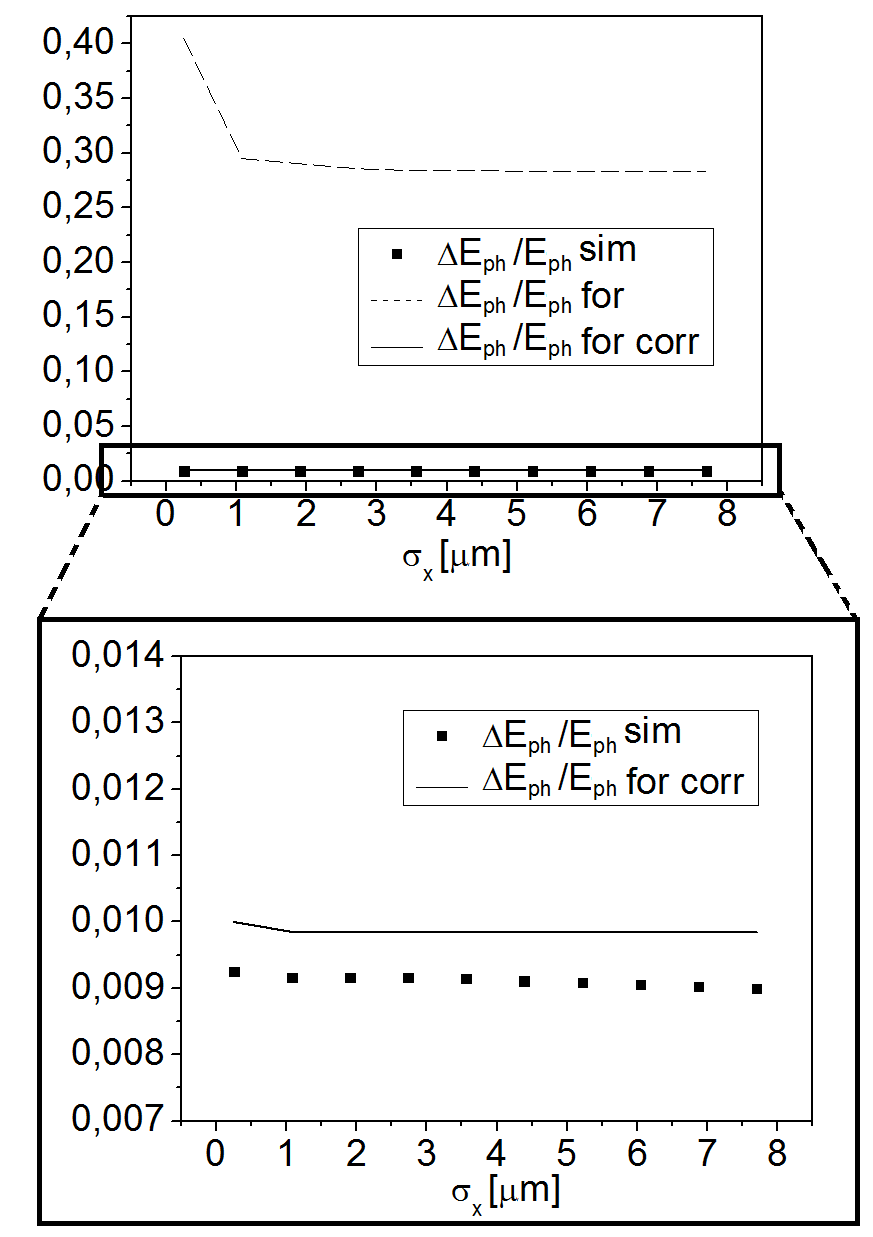}
	\caption{Case F: $\Delta E_{ph}/E_{ph}$ value from CAIN simulation (squares) vs formula (\ref{f1}) without (dashed line) and with (solid line) $X$ correction for different $\sigma_x$ ($\mu$m) values calculated at $\theta=500$ $\mu$rad and $\epsilon_n=0.082$ $\mu$m\ rad.}\label{sighigh}
\end{figure}

\begin{figure}[htbp]\centering
	\includegraphics [scale=0.28] {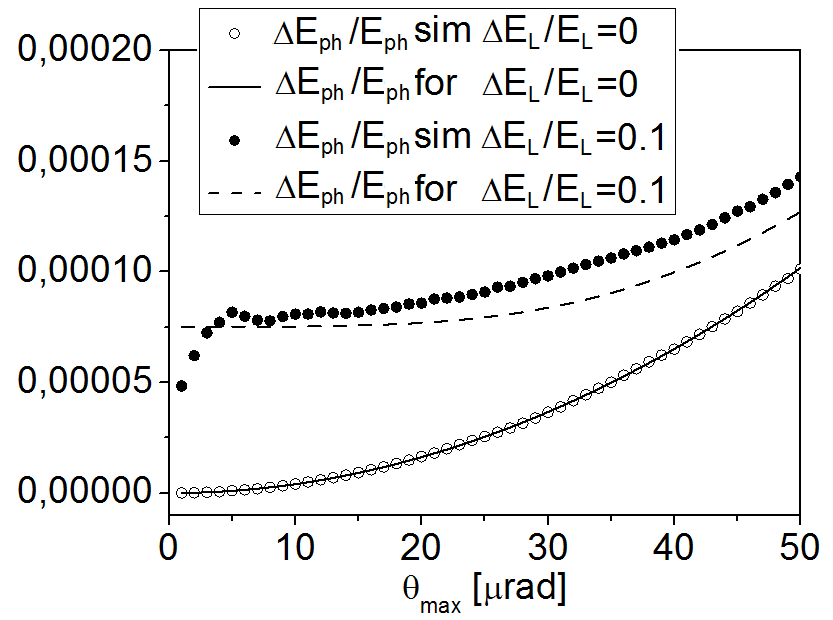}
	\caption{Case F: $\Delta E_{ph}/E_{ph}$ value from CMCC simulation (circles) vs formula (\ref{f1}) (lines) as a function of $\theta_{max}$ ($\mu$rad), for $\Delta E_{L}/E_{L}=0$ and $\Delta E_{L}/E_{L}=0.1$.}\label{varnu0}
\end{figure}

\begin{figure}[htbp]\centering
	\includegraphics [scale=0.28] {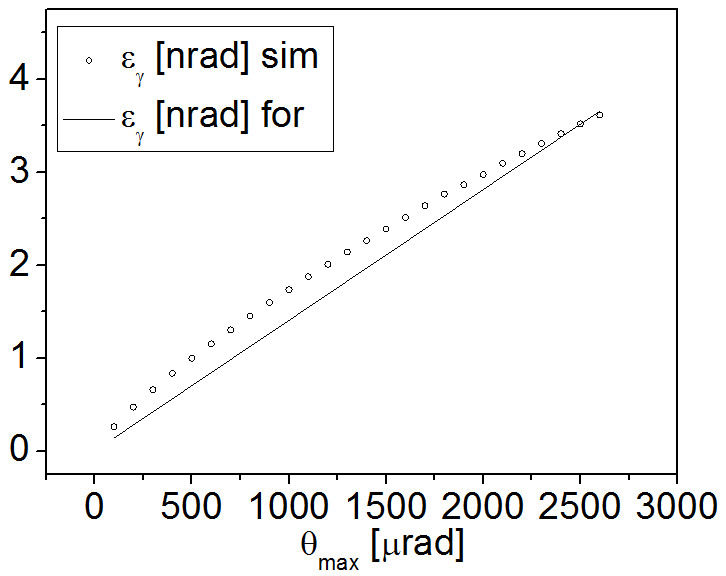}
	\caption{Case F: $\epsilon_{\gamma}$ value from CAIN simulation (circles) vs formula (\ref{ag}) (solid line) [nrad] as a function of $\theta_{max}$ ($\mu$rad).}\label{emittgammahhigh}
\end{figure}

\noindent The most unusual regime is the one at very large recoil: here the $X$ contribution is fundamental in the $\Delta E_{ph}/E_{ph}$ calculation (Fig. \ref{low}). The energy spectrum is peaked around the incoming electron beam energy (Figs. \ref{highvs}, \ref{caseD}). Most of the incoming electron energy is transferred to the photons which are emitted on a wider rms $\theta$ with respect to the low recoil case. This large recoil regime shows some very peculiar characteristics not present in the other cases: the bandwidth value becomes insensitive to some of the incoming beams features such as the electron beam emittance and the laser bandwidth. As presented in Fig. \ref{emitthigh}, even a big increase in the $\epsilon_n $ value corresponds to a really modest increase of the emitted radiation bandwidth. As shown in Fig. \ref{sighigh}, the variation of $\sigma_x$ does not affect the bandwidth value either. In the same way a huge $10\%$ increase of $\Delta E_{L}/E_{L}$ leads to a small $7\cdot 10^{-5}$  $\Delta E_{ph}/E_{ph}$ broadening (Fig. \ref{varnu0}).\\ 
\noindent The emittance of the scattered photon beam described by eq. (\ref{emit}) is in agreement with the simulations (see Fig. \ref{emittgammahhigh}), therefore also the values of peak and average brilliance predicted by formulas (\ref{b}), (\ref{ba}) are consistent with the Monte Carlo results.

\clearpage
\section{Conclusions}
We derived simple analytical formulas able to predict with quite good accuracy the characteristics of 6D phase space distributions of the X/$\gamma$ photon beams generated by back-scattering in Inverse Compton Sources. The formulas are strictly valid in the linear regime. Since most of ICSs in operation, under construction and design, to be  implemented as user facilities, are meant to be operated in linear or weak non-linear regime, the set of formulas here presented constitutes a useful scheme of guidelines to help ICSs designers with quick predictions of their anticipated performances in terms of flux, bandwidth, spectral density, emittance and brilliance of the photon beam.\\
In particular the formulas, derived by developing the kinematics in the center of mass reference system of the electron-photon collision, are valid for any value of the electron recoil, covering the whole range of energy of colliding electrons (MeV to multi-GeV) and incident photons (from optical laser pulses to multi-keV FEL's), with the only restriction of relativistic electrons and incident photons with energy much smaller than the electron one. Under this respect, such a generalization to any value of the electron recoil represents an upgrade with respect to previous works.\\
Two somewhat new findings \cite{hajima} that occur at large recoil, $X\gg 1$, have been theoretically explained throughout this paper:
\begin{itemize}
\item[i.]a suppression of the bandwidth dependence on the electron beam emittance;
\item[ii.]a much weaker dependence of the bandwidth on the frequency spread of the incident photon beam.
\end{itemize}
These effects are basically negligible in the usual regime of ICSs, not only the so-called Thomson regime for X-rays used for radiological imaging, but even those used for nuclear physics/photonics with MeV-class photon beams. Only when higher energy incident photon beams are considered, like for instance with X-ray FEL's, the effects come to play with full deployment, and can be exploited in a strategic way to optimize the design of recoil dominated ICSs. Allowing for instance an over focusing of the electron beam to maximize luminosity without spoiling the photon beam bandwidth, which is not allowed in low recoil ICSs due to the dependence of bandwidth on the electron beam emittance (i.e. the rms electron beam transverse momentum). Also the use of a broad-band incident photon beam would be possible in large recoil ICSs, without spoiling the photon bandwidth, thanks to the large suppression applied by the large recoil factor.\\
\noindent The summary of formulas is reported in the following:\\

	\begin{widetext}
		\begin{equation*}
		X=\frac{4\,E_e\,E_{L}}{M_e^2} \hspace{2cm}	\gamma_{CM}=\frac{\gamma}{\sqrt{1+X}} \hspace{2cm}
			\Psi=\gamma_{CM}\theta_{max} 
			\end{equation*}	
		\begin{equation*}
		\overline{P}=\frac{\sqrt{2}\epsilon_n}{\sigma_x\sqrt{1+X}}=\frac{\gamma_{CM}}{\gamma}\frac{\sqrt{2}\epsilon_n}{\sigma_x} \hspace{2cm}
			E_{ph}=4\,\gamma_{CM}^{2}\,E_{L}\left(1-\gamma_{CM}^{2}\,\theta^{2}\right)
		\end{equation*}
			\begin{equation*}
			\frac{\Delta E_{ph}}{E_{ph}}\simeq \sqrt{\left[\frac{\Psi^2/\sqrt{12}}{1+\Psi^2}+\frac{\overline{P}^2}{1+\sqrt{12}\,\overline{P}^2}\right]^2+\left[\left(\frac{2+X}{1+X}\right)\frac{\Delta\gamma}{\gamma}\right]^2+\left(\frac{1}{1+X}\frac{\Delta E_L}{E_L}\right)^{2}+
				\left(\frac{M^{2}\lambda_{0}}{2\pi w_{0}}\right)^{4}+\left(\frac{a_{0}^{2}/3}{1+a_{0}^{2}/2}\right)^{2}}		
			\end{equation*} 
			\begin{equation*}
			\mathcal{N}^{\Psi}=6.25\cdot 10^8\frac{ U_L(J) \, Q(pC) \, r}{E_{L}(eV)\left(\sigma_{x}^2(\mu m)+\sigma_L^2(\mu m)\right)}
			\frac{\left(1+\sqrt[3]{X}\Psi^2/3\right)\Psi^2}{\left(1+(1+X/2)\Psi^2\right)(1+ \Psi^2)}
			\hspace{2cm}
			S=\frac{\mathcal{N}^{\Psi}}{\sqrt{2\,\pi}\, 4\,E_L\gamma_{CM}^2 \,\frac{\Delta E_{ph}}{E_{ph}}}
			\end{equation*}
			\end{widetext}
		
The analysis reported here is valid for unpolarised beams and a perfectly head-on collision. The generalisation to small non-zero collision angle is done by correcting the recoil factor to \\
\begin{equation}
   	X=\frac{2\,E_e\,E_{L}}{M_e^2}\,(1+\cos\alpha)\\
\end{equation}
where $\alpha$ is the (small) collision angle. The correction to luminosity due to non-zero collision angle can be approximately taken care by multiplying the luminosity by the correction factor \cite{lum}
\begin{equation}
	\delta_{\alpha}=\frac{1}{\sqrt{1+\frac{\alpha^2(\sigma_z^2+c^2\sigma_t^2)}{4(\sigma_x^2+\sigma_L^2)}}}.
\end{equation}

\end{document}